\documentclass[12pt,preprint]{aastex}

\newcommand{\geff}{\mbox{$g_{\rm eff}$}}
\newcommand{\Bz}{\mbox{$B_{\rm z}$}}

\begin{document}

\title{\bf Spectropolarimetry of the Classical T Tauri Star TW Hydrae}

\author{Hao Yang}
\affil{Department of Physics \& Astronomy, Rice University, 6100 Main St.
       MS-108, Houston, TX 77005}
\email{haoyang@rice.edu}

\author{Christopher M. Johns-Krull\altaffilmark{1}}
\affil{Department of Physics \& Astronomy, Rice University, 6100 Main St.
       MS-108, Houston, TX 77005}
\email{cmj@rice.edu}

\author{Jeff A. Valenti\altaffilmark{1}}
\affil{Space Telescope Science Institute, 3700 San Martin Dr., Baltimore, MD
       21210}
\email{valenti@stsci.edu}

\altaffiltext{1}{Visiting Astronomer, McDonald Observatory, operated
by The University of Texas at Austin}

\begin{abstract}

We present high resolution ($R \approx 60,000$) circular
spectropolarimetry of the classical T Tauri star TW Hydrae. We
analyze $12$ photospheric absorption lines and measure the net
longitudinal magnetic field for 6 consecutive nights. While no net
polarization is detected the first five nights, a significant
photospheric field of $B_{\rm{z}} = 149 \pm 33$ G is found on the
sixth night. To rule out spurious instrumental polarization, we
apply the same analysis technique to several non-magnetic telluric
lines, detecting no significant polarization. We further demonstrate
the reality of this field detection by showing that the splitting
between right and left polarized components in these 12 photospheric
lines shows a linear trend with Land\'e $g$-factor times wavelength
squared, as predicted by the Zeeman effect. However, this
longitudinal field detection is still much lower than that which
would result if a pure dipole magnetic geometry is responsible for
the mean magnetic field strength of $2.6$ kG previously reported for
TW Hya. We also detect strong circular polarization in the
\ion{He}{1} $5876$ \AA\ and the \ion{Ca}{2} $8498$ \AA\ emission
lines, indicating a strong field in the line formation region of
these features. The polarization of the \ion{Ca}{2} line is
substantially weaker than that of the \ion{He}{1} line, which we
interpret as due to a larger contribution to the \ion{Ca}{2} line
from chromospheric emission in which the polarization signals
cancel.  However, the presence of polarization in the \ion{Ca}{2}
line indicates that accretion shocks on Classical T Tauri stars do
produce narrow emission features in the infrared triplet lines of
Calcium.

\end{abstract}

\keywords{ Stars: spectropolarimetry --- stars: magnetic fields ---
stars: pre--main sequence --- stars: individual (TW Hya) }

\section{Introduction}

T Tauri stars (TTSs) are newly formed low-mass stars that have
recently become visible at optical wavelengths. These young, roughly
solar mass stars are still contracting along pre-main sequence
evolutionary tracks in the H-R diagram. It is generally believed
that classical T Tauri stars (CTTSs) are still surrounded by disks
of material that are undergoing accretion onto the central star,
producing excess emission in both lines and continuum at multiple
wavelengths. Magnetospheric accretion models are the most popular
description of the accretion process. Strong, stellar magnetic
fields are believed to regulate the accretion and confine disk
material to flow onto the stellar surface along the field lines
(e.g., Camenzind 1990; K\"onigl 1991; Cameron \& Campbell 1993; Shu
et al. 1994). These models generally assume the magnetic structure
of TTSs to be dipolar and require magnetic field strengths which
vary over a wide range of values, with the field on some stars as
high as several kilogauss (see Johns--Krull, Valenti \& Koresko
1999b). Such high field strengths should be measurable by utilizing
the most magnetically sensitive diagnostics.

On the other hand, direct magnetic field measurements are difficult,
since T Tauri stars are relatively faint and display various
spectral peculiarities. The most successful approach for measuring
fields on late--type stars in general has been to measure Zeeman
broadening of spectral lines in unpolarized light (e.g., Robinson
1980; Saar 1988; Valenti, Marcy \& Basri 1995; Johns--Krull \&
Valenti 1996; Johns--Krull et al. 1999b; Johns--Krull, Valenti, \&
Saar 2004). This method is more efficient at infrared wavelengths
thanks to the $\lambda^{2}$ dependence of Zeeman broadening,
compared to the $\lambda^1$ dependence of Doppler broadening. While
a sensitive measure of field strength, Zeeman broadening
measurements give little information on the magnetic field geometry.

Another direct method for measuring magnetic fields is to detect net
circular polarization in Zeeman-sensitive lines. Generally, Zeeman
$\sigma$ components are elliptically polarized and the components of
opposite helicity are split to either side of the nominal line
wavelength. A net longitudinal component of the magnetic field makes
components of Zeeman-sensitive lines distinguishable through a
right-circular polarizer (RCP) and left circular polarizer (LCP).
The separation between the line observed in RCP and LCP light is
$$\Delta\lambda = 2{e \over 4\pi m_ec^2} \lambda^2 g_{\rm{eff}} B_{\rm{z}}
                = 9.34 \times 10^{-7} \lambda^2 g_{\rm{eff}} B_{\rm{z}} \,\,\,\,\,\,
                  \rm m\mbox{\AA}\,\eqno(1)$$
where $g_{\rm{eff}}$ is the effective Land\'e $g$-factor of the
transition, $B_{\rm{z}}$ is the strength of the mean longitudinal
magnetic field in kilogauss, and $\lambda$ is the wavelength of the
transition in Angstroms (see Mathys 1988, 1991). The weights for
individual $\pi$ and $\sigma$ components in the definition of
$g_{\rm eff}$ assume an optically thin medium, so eq.[1] is only
approximately true in our case of moderately strong photospheric
lines. Previously, Johns--Krull et al. (1999a) did not detect
polarization in the photospheric lines of the CTTS BP Tau, setting a
$3\sigma$ upper limit on \Bz\, of $\pm 200$ G.  Smirnov et al.
(2003) report a longitudinal magnetic field of $B_{\rm{z}} \sim 150
\pm 50$ G on the CTTS T Tau, which is very close to their detection
limit, while their subsequent observation of T Tau (Smirnov et al.
2004) did not detect a significant field. In an effort to confirm
the original Smirnov et al. (2003) detection, Daou et al. (2006)
measure a mean longitudinal field of \Bz$\, = 12\,\pm \,35$ G on T
Tau.  Daou et al. (2006) use upper limits on \Bz\, on multiple
nights along with the mean field strength detected on T Tau of $\bar
B \sim 2.4\ $kG (Guenther et al. 1999; Johns--Krull et al. 2001) to
seriously question the assumed dipole field geometry (see also
Valenti \& Johns--Krull 2004).

     On the other hand, Johns--Krull et al. (1999a) discovered
net polarization in the \ion{He}{1} $5876$ \AA\ emission line on the
CTTS BP Tau, indicating a net longitudinal magnetic field of $2.46
\pm 0.12$ kG in the line formation region. This \ion{He}{1} emission
line is believed to be produced, at least partially, in the shock
region formed where disk material accretes onto the stellar surface
(Hartmann, Hewett \& Calvet 1994; Edwards et al. 1994). Circular
polarization in the \ion{He}{1} line has now been observed in
several CTTSs (Valenti \& Johns--Krull 2004; Symington et al. 2005).
These observations suggest that accretion onto CTTSs is indeed
controlled by a strong stellar magnetic field.

While substantial observational evidence indicates strong fields on
the surface of TTSs, the origin of the surface magnetic fields on
TTSs are not clear.  Interface dynamo models (e.g., Parker 1993)
which are applied to solar-type main-sequence stars probably do not
apply to TTSs since their internal structure is significantly
different from that of the Sun. Feigelson et al. (2003) summarize
several theoretical considerations for the origin of TTS magnetic
fields. One possibility is that a distributed dynamo due to
turbulent convection could operate in TTSs and generate small-scale
magnetic fields. These fields could also be amplified by
differential rotation throughout the convective zone (Durney et al.
1993; Kitchatinov \& R\"udiger 1999; K\"uker \& Stix 2001), though
surface differential rotation is weak or absent on TTSs (Johns-Krull
1996). It is also possible that T Tauri stars simply maintain a
``fossil" magnetic field throughout the star formation process and
do not have significant field contributions from dynamo processes
(Tayler 1987; Mestel 1999; Moss 2003).  More observational studies
are needed to put further constraints on theories of the origin and
evolution of magnetic fields on pre-main sequence stars.

In order to gain further insight into the magnetic properties of
young stars, we present an analysis of high resolution
spectropolarimetry of the CTTS TW Hya, a K7Ve(Li) star (Herbig 1978)
located at a distance of $56 \pm 7$ pc (Wichmann et al. 1998).
Imaging in multiple wavelengths (e.g., Krist et al. 2000; Weinberger
et al. 2002; Qi et al. 2004) indicates a nearly face-on disk around
TW Hya that is optically thick in the visible, near-IR,
sub-millimeter and millimeter wavelengths. Herczeg et al. (2004)
estimate an accretion rate of $2 \times 10^{-9}$\ ${\rm M}_\odot\
{\rm yr}^{-1}$ from its strong optical and near-UV excess emission.
Previously, we have modeled magnetic broadening of four K band
\ion{Ti}{1} lines observed in unpolarized light to measure a mean
magnetic field strength of $\bar B = 2.6 \pm 0.2$ kG on TW Hya
(Yang, Johns--Krull \& Valenti 2005). Here, we use a time series of
spectropolarimetric observations to study the geometry of the
stellar magnetic field. The remainder of this paper is organized as
follows. In \S\ 2 we describe our observations and data reduction
procedures. In \S\ 3 we present the measurements of the longitudinal
field in the photosphere as well as in the accretion regions.
H$\alpha$ line profiles are also presented in \S\ 3. Finally, a
discussion of our results is presented in \S\ 4.

\section{Observations and Data Reduction}

We obtained spectra using a Zeeman analyzer (ZA) system on the 2.7 m
Harlan J. Smith Telescope at McDonald Observatory on April 21-26,
1999. This system has been described by Vogt et al. (1980), with
subsequent modifications described by Johns--Krull et al. (1999a).
The ZA splits stellar light coming to a focus on the slit into two
parallel beams that create two separate stellar images on the
spectrograph slit. One beam contains approximately half the
unpolarized light and any RCP light, while the other beam contains
the remainder of the unpolarized light and any LCP light.

The ZA was used with the 2-d coud\'e cross-dispersed  echelle
spectrometer (Tull et al. 1995). This spectrometer provides a
2-pixel spectral resolution of $R \equiv \lambda/\delta\lambda
\approx 60,000$ and enough space between the orders to interleave
simultaneously stellar spectra of both circular polarization states.
To reduce spurious linear polarization induced by the coud\'e mirror
train, the ZA control computer automatically updates the retardance
of a Babinet-Soleil phase compensator (PC) fixed to the front of the
ZA. The phase compensation is continually changed, as telescope
orientation changes throughout an observation. Each night, 2
exposures of TW Hya were obtained. Before the second exposure, an
achromatic 1/2-wave plate (manufactured by Special Optics, model No.
8-9012-1/2) was inserted in front of the ZA+PC in order to switch
the sense of circular polarization recorded in the two interleaved
spectra. Analyzing and averaging the results of this pair of
exposures reduces potential sources of systematic/instrumental error
in the measurements.  All spectra were reduced using an
echelle-reduction package developed by Valenti (1994) and described
more fully in Hinkle et al. (2000). Wavelength solutions are
determined from spectra of a Thorium-Argon lamp by performing a
two-dimensional fit to the positions of lines on the detector as a
function of $n$ and $n\lambda$, where $n$ is the echelle order.
Table 1 summarizes the observations discussed here.

\section{Analysis}

\subsection{Photospheric $B_{\rm{z}}$}

We use 12 photospheric absorption lines that have relatively large
Land\'e $g$-factors to measure the longitudinal field, $B_{\rm z}$,
on TW Hya. These lines are also relatively strong and unblended, and
not significantly affected by telluric absorption. The species we
use, as well as their wavelengths and Land\'e $g$-factors, are
listed in the first three columns of Table \ref{linedata}.

Our analysis technique is as follows. We cross-correlate the LCP and
RCP line profiles and measure the wavelength separation between the
two spectra for each line. Using eq[1], we convert the measured
wavelength separation for each line into a longitudinal magnetic
field. We then use a Mont\'e Carlo analysis to estimate the
uncertainties in our measured wavelength separations. First we fit a
Gaussian curve to the observed intensity profile, which is the sum
of the LCP and RCP spectra for each night. By adding noise
comparable to that in our observations to the Gaussian curve, we
construct a pair of synthetic observations (one represents the LCP
and the other the RCP component). Then we analyze these profiles in
the same way as we handle the actual observations and obtain a line
separation which is then translated into magnetic field strength
using eq.[1]. For each individual spectral line on each night, we
execute the Mont\'e Carlo process above 100 times and measure the
apparent shift (we have tried shifts of 0 and 0.5 pixels) between
the synthetic LCP and RCP spectra. We then adopt the standard
deviation of the corresponding Mont\'e Carlo results as the
uncertainties in our measurement of $B_{\rm{z}}$. The results for
each photospheric line for each of the six nights are listed in the
last six columns of Table \ref{linedata}. The weighted mean field
values and their uncertainties for the six nights are listed in
Table 1. The night-to-night variation is plotted at the bottom panel
of Figure \ref{tseries}.

The mean longitudinal field value on April 26, 1999 is
$149\,\pm\,33$ Gauss, well over the $3\sigma$ limit. If this field
strength and uncertainty estimate are accurate, it represents the
largest magnitude of $B_{\rm z}$ detected at a significant level on
a low mass pre-main sequence star, and one of the highest $B_{\rm
z}$ values detected for any low mass star (see Donati et al. 2006).
To rule out spurious instrumental polarization, we apply the same
technique to analyze six magnetically insensitive telluric
absorption lines and find the observed wavelength separations
between the LCP and RCP spectra of these lines. The telluric lines
are narrower than the stellar photospheric lines, allowing
wavelength separation to be measured more precisely, yielding
smaller uncertainties in \Bz, and placing tighter limits on spurious
instrumental polarization. In order to translate the observed
wavelength separations into magnetic field strengths for comparison,
we assign a \geff\ for the telluric lines of 0.93, which is
determined from the weighted mean value of $\lambda^2$\geff\ for the
12 photospheric lines, where the weights are uncertainties in the
photospheric field measurements. The field estimates from the
telluric lines are given in Table \ref{linedata} for each night. The
recovered field strengths for the telluric lines are all consistent
with no magnetic field (as they should be) to within the errors,
which are typically $\sim 28$ G ($1\sigma$).

\begin{figure}[ht]
  \begin{center}
  \includegraphics[scale=0.7]{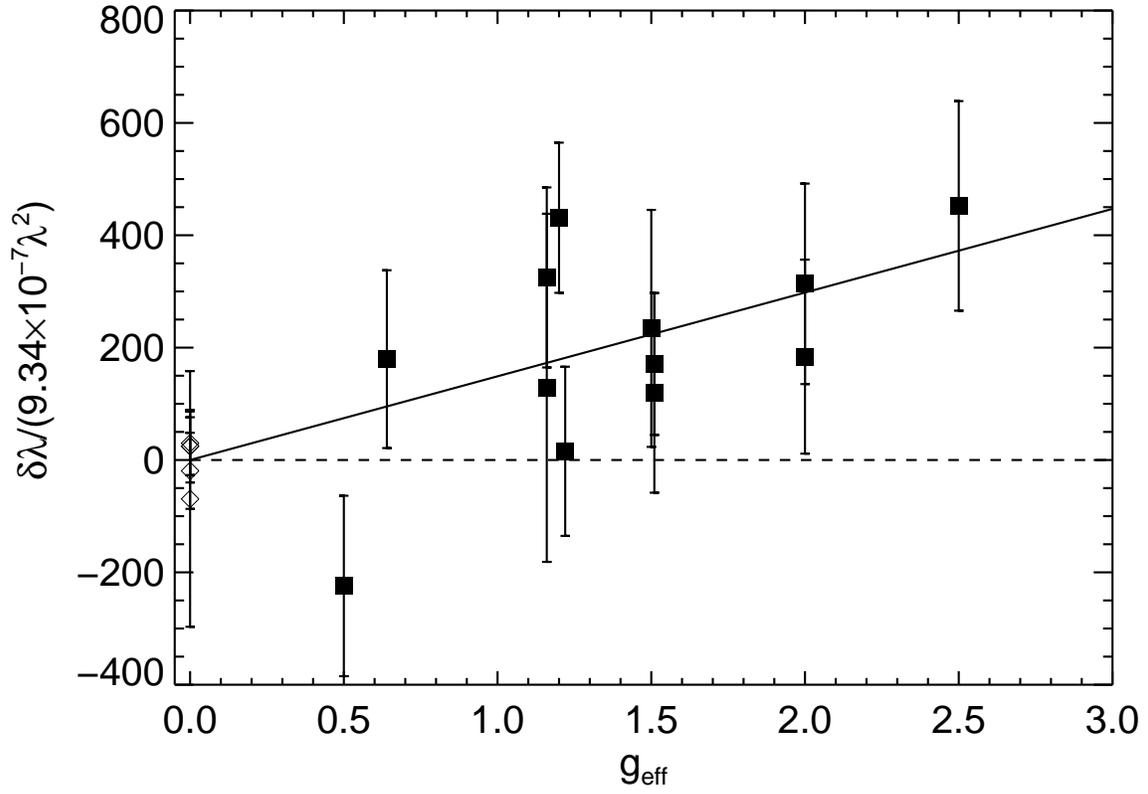}
  \end{center}
  \caption{Plot of $\delta\lambda/(9.34\times10^{-7}\lambda^2)$ against
  Land\'e $g$-factors for the data from April 26, 1999. The solid
  line marks where $B_{\rm z}$ is equal to 149 G, the measured mean value.
  (Filled squares: photospheric lines; diamonds: telluric lines.)}
\label{sixth}
\end{figure}

Another way to confirm the reality of the photospheric $B_{\rm z}$
measurement on the last night of observation is to look for a
correlation between the measured wavelength shift and the \geff\ of
each line. This is perhaps the best way to establish the magnetic
origin of the shifts, and hence rule out any potential instrumental
effects unaccounted for. We can rewrite eq.[1] as follows:
$${\Delta\lambda \over 9.34 \times 10^{-7} \lambda^2}
                = B_{\rm{z}} g_{\rm{eff}}  \,\,\,\,\,\, \eqno(2)$$
In Figure \ref{sixth}, we plot the left hand side of eq.[2] against
 \geff. The solid line marks the expected relationship for $B_{\rm
z} = 149$ G. The wavelength separations of the telluric lines
(hollow diamonds in Figure \ref{sixth}) are all found to be close to
zero as they should be, since the molecular lines have negligible
\geff\ and form in the weakly magnetized atmosphere of the Earth.
Using all the data points in Figure \ref{sixth}, the reduced
$\chi^2$, $\chi^2_{\rm r}$, for the $B_{\rm z} = 149$ G line is
$0.68$, corresponding to an $83\%$ chance of being an acceptable
model, while a best-fit horizontal line, indicative of an
instrumental offset, yields $\chi^2_{\rm r} = 1.91$, corresponding
to less than a $2\%$ chance of being an acceptable model. The
positive correlation shown in Figure \ref{sixth} and the detailed
statistical tests give confidence that the measured wavelength
separations are indeed magnetic in origin, so that we do detect a
rather strong longitudinal field on TW Hya above the $4\sigma$
limit.

Examination of Table 1 or the bottom panel of Figure 3 shows that
while we find a value of \Bz\ larger than the $3\sigma$ measurement
uncertainties only once, all measurements are systematically
positive and agree with each other within the uncertainties.  Given
that TW Hya has an inclination close to 0$^\circ$, little rotational
modulation is expected (though see \S 3.2).  Taking the weighted
mean of all 6 nights data gives a value of \Bz$ = 90 \pm 17$ G.

\subsection{Magnetic Fields in the Emission Line Region}

Significant polarization in the \ion{He}{1} $5876$ \AA\ emission
line has been detected on several CTTSs. Johns-Krull et al. (1999a)
first discovered this polarization and found $B_{\rm z} = 2.46 \pm
0.12$ kG in the \ion{He}{1} line formation region of BP Tau. Valenti
\& Johns-Krull (2004) found \ion{He}{1} polarization in four CTTSs:
AA Tau, BP Tau, DF Tau, and DK Tau. Symington et al. (2005) also
detect \ion{He}{1} polarization at greater than the $3\sigma$ level
in three stars (BP Tau, DF Tau, and DN Tau) in their survey of seven
CTTSs. While this \ion{He}{1} line can form weakly in emission in
naked TTSs (NTTSs) which are believed to lack close circumstellar
disks and significant accretion, the strong \ion{He}{1} emission of
CTTSs is thought to form in the accretion shock region where disk
material hits the stellar surface (e.g., Edwards et al. 1994;
Hartmann, Hewett \& Calvet 1994).

\begin{figure}[ht]
  \begin{center}
  \includegraphics[scale=0.65]{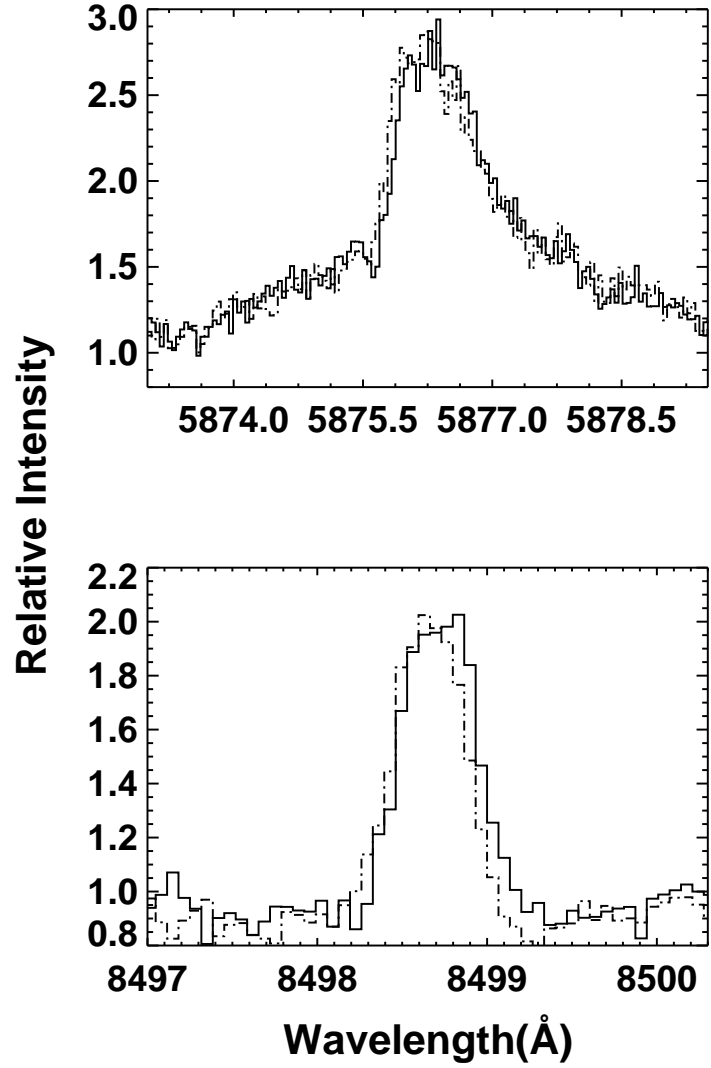}
  \end{center}

  \caption{ Observed LCP(solid line) and RCP(dash-dotted line)
  spectra of \ion{He}{1} $\lambda5876$ line and \ion{Ca}{2} $\lambda8498$ line on
   April 26, 1999.}

 \label{heca}
\end{figure}

\begin{figure}[ht]
  \begin{center}
  \includegraphics[scale=0.65]{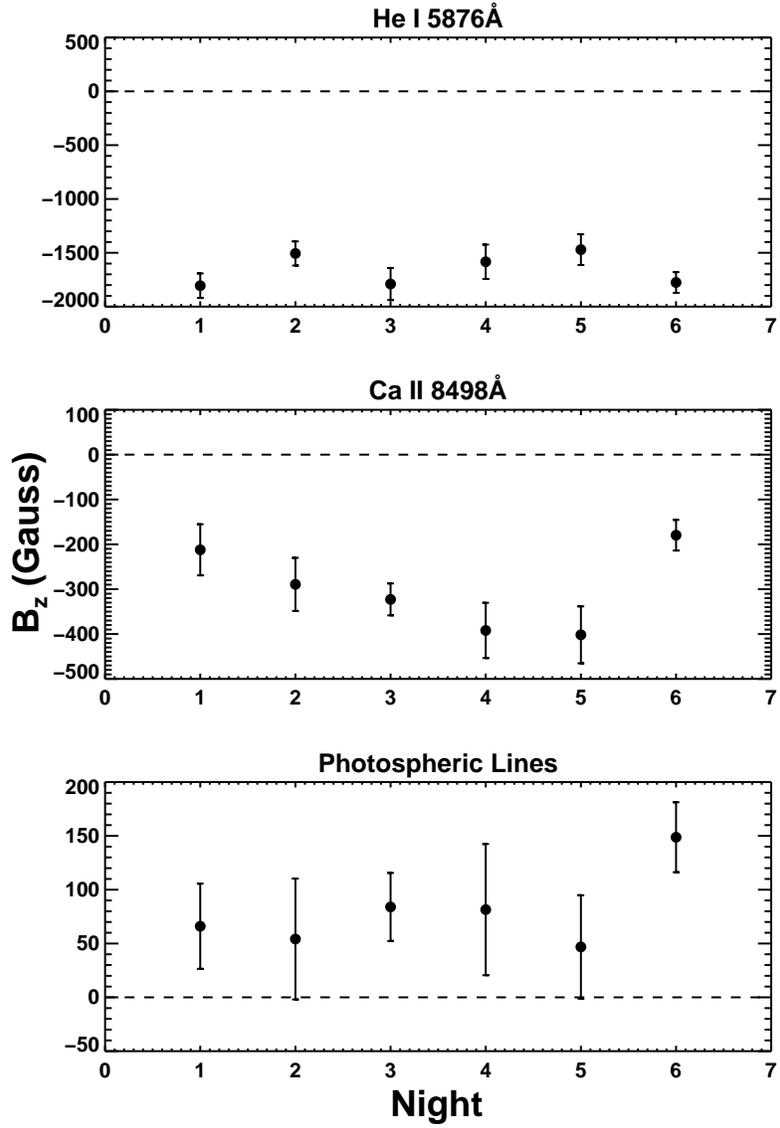}
  \end{center}
  \caption{Time series of $B_{\rm{z}}$ values measured from different spectral
  lines.(Top Panel: field values measured from \ion{He}{1} $\lambda5876$ line; Middle Panel:
  measured from \ion{Ca}{2} $\lambda8498$ line; Bottom Panel: measured from 12
  photospheric lines.)}
\label{tseries}
\end{figure}

Here we analyze our observations of TW Hya and find strong circular
polarization in the \ion{He}{1} $5876$ \AA\ emission line as well. The same
analysis technique used for the photospheric lines is applied. This
\ion{He}{1} line is a multiplet.  The observed shift
between the line observed in different circular polarization states
described below is larger than the spacing between most of the
multiplet members that make up this feature.  As a result, the
magnetic splitting is best described by the Paschen-Bach effect,
hence \geff\ = $1.0$ for the line. [Even in the weak field limit of
Zeeman broadening, LS coupling gives \geff\ = $1.11$ (Johns--Krull
et al. 1999a), so the choice of treatment makes only a small
difference on the resulting field values.]  The measurements for
each night are listed in Table \ref{linedata} and the weighted mean of
the net longitudinal field for all the nights together is
$-1673\pm50$ G. A pair of representative LCP and RCP spectra is
shown in top panel of Figure \ref{heca}.  In addition to the 5876 \AA\
line of \ion{He}{1}, our spectrometer setting also contains the 6678 \AA\
line of \ion{He}{1}.  This line is the singlet counterpart to the 5876 \AA\
line and has \geff\ = 1.0.  Analysis of the 6678 \AA\ line gives magnetic
field values in the \ion{He}{1} line formation fully consistent with the
values from the 5876 \AA\ line to within the measurement uncertainties
(which are about a factor of 2 larger for the 6678 \AA\ line since this
line is weaker than the 5876 \AA\ line).

We also detect significant polarization in the \ion{Ca}{2} $8498$
 \AA\ emission line as shown in the bottom panel of Figure
\ref{heca}. Nightly measurements are given in Table \ref{linedata}
and we measure a weighted mean $B_{\rm z} = -276 \pm 19$ G for all
the nights together.  We expect that the other members of the
\ion{Ca}{2} infrared triplet (IRT) show similar levels of
polarization, but they fall in the gaps in our spectral coverage and
were not observed. The origin of the narrow IRT emission lines seen
in many CTTSs (including TW Hya) is somewhat debated and is
discussed further in \S4. Our detection of polarization in this line
with the same polarity as the \ion{He}{1} line suggests that the
accretion shock contributes some emission to this component of the
IRT. Our measured \Bz\ for the \ion{Ca}{2} line is 16\% of our
measured \Bz\ for the \ion{He}{1} line, which suggests at least 16\% of the
\ion{Ca}{2} emission comes from accretion regions with highly
ordered magnetic fields. This is a lower limit because the lower
\Bz\ for \ion{Ca}{2} could also signify a larger contribution from
regions of lower field strength (e.g., the accretion column) or
mixed polarity (e.g., a radiatively heated sheath around the
accretion footpoint as proposed by Batalha et al. 1996).
Non-shock contribution to the \ion{He}{1} line
could lower our estimate of 16\%, but in the case of TW Hya the
effect is almost negligible (see more on this issue at the end of
\S4).

The time series of the measured field values from the \ion{He}{1}
and \ion{Ca}{2} lines, along with that from the photosphere, are
plotted in Figure \ref{tseries}. The night-to-night variation of the
polarization in the \ion{He}{1} line shows a hint of periodicity. We
adopt a rotation period $P = 2.2$ days for TW Hya from Makkaden
(1998) and fit a sine wave to the measured field values. The fit has
$\chi^2=1.2$ for 3 degrees of freedom, indicating an $76\%$
probability of being an acceptable model. However, we also fit a
straight line to our data (representing a model with no variability)
and find $\chi^2 = 6.30$ for 5 degrees of freedom, which has a
significant $28\%$ probability of representing an acceptable model.
Both fits are shown in Figure 4. The \ion{Ca}{2} and photospheric
lines do not show the same indications for periodicity; however,
their relative uncertainties are much larger. Since our analysis is
limited to six data points, the evidence for rotational modulation
in the \ion{He}{1} line is suggestive, but remains inconclusive.

\subsection{H$\alpha$ Line Profiles}

In Figure \ref{halpha}, we plot the H$\alpha$ profiles from all six
nights. The intensity of the H$\alpha$ line varies and generally
decreases with time, suggesting a possible decrease in the accretion
rate on later nights. A variable blue shifted absorption component
indicative of the wind from TW Hya also decreases with time over
these six nights. Other than these general trends in the H$\alpha$
line over these 6 nights, nothing dramatic occurs on the last night
when the photospheric $B_{\rm z}$ takes on its largest value.

We looked for circular polarization in the H$\alpha$ emission line
as well.  This line is very strong which aids the detection of weak
line shifts between the RCP and LCP light.  However, since this data
was obtained with an echelle spectrometer which has a strong blaze
function, continuum normalization under H$\alpha$ is not trivial.
Small differences in the continuum normalization of the RCP and LCP
profiles can produce spurious wavelength shifts.  We attempt to
quantify this by repeating the analysis of the H$\alpha$ profile
using somewhat different regions to fit the continuum, as well as
using polynomials of order 4, 5, and 6 in the continuum
normalization process.  For a given continuum fit, the strength of
the H$\alpha$ line results in a \Bz\ uncertainty of 6 G resulting
from the signal-to-noise in the spectra.  On the other hand, our
different continuum fits on the same observed spectrum yielded field
measurements different by as much as 46 G in some cases.  Therefore,
we conservatively adopt 46 G as the uncertainty in our H$\alpha$
\Bz\ measurements.  We did not detect a value of \Bz\ significantly
above this level on any of the nights we observed TW Hya, so we put
a conservative upper limit of \Bz\ in the H$\alpha$ line formation
region of 138 G.

\section{Discussion}

Imaging of the circumstellar disk around TW Hya in the infrared
(Krist et al. 2000; Trilling et al. 2001; Weinberger et al. 2002),
millimeter (Wilner et al. 2000) and sub-millimeter (Qi et al. 2004)
wavelengths all suggest that the inclination of the disk is close to
$0^\circ$. Alencar \& Batalha (2002) derived an inclination of
$18^\circ \pm 10^\circ$ from emission line profile analysis. Lawson
and Crause (2005) derive $i \sim 16^\circ$ from photometric
monitoring. The night-to-night variation in our measurements of the
longitudinal field in the \ion{He}{1} line is small, only about
$10\%$ of the mean value, which is also consistent with a low disk
inclination angle of TW Hya.  Such a geometry allows a strong test
whether the magnetic field on TW Hya is primarily a dipole field
with the magnetic axis aligned with the rotation axis. Yang et al.
(2005) find the mean magnetic field strength in the photosphere of
TW Hya to be $2.6 \pm 0.2$ kG from infrared (IR) Zeeman broadening
measurements. If we follow Alencar \& Batalha (2002) and assume an
inclination angle between $8^\circ$ and $28^\circ$, and if the
magnetic dipole axis is aligned with the rotation axis, the $2.6$ kG
mean field would predict a mean line of sight field in the
photosphere of \Bz \ $ = 0.97 - 1.05$ kG. This is much higher than
our maximum measured value \Bz$ = 149 \pm 33$ G, or our weighted
mean from all nights of \Bz$= 90 \pm 17$ G. One possibility to
explain the low longitudinal magnetic field values yet retain a
dipole field geometry is to assume the magnetic axis is highly
inclined from the rotation axis (i.e. has high obliquity, $\beta$).
For example, Krist et al. (2000) use the model of Mahdavi \& Kenyon
(1998) and conclude that the photometric variability of TW Hya
observed by Mekkaden (1998) can be explained if $i < 10^\circ$ and
the field is a dipole with $\beta > 55^\circ$. If we take $i =
10^\circ$ and $\beta = 55^\circ$, the 2.6 kG mean field on TW Hya
still implies a photospheric \Bz$ = 537$ G, again well above our
upper limits for \Bz\ in the photosphere. If the photospheric field
is globally dipolar, then $\beta$ must be significantly larger than
$55^\circ$. If we ask what is required of a dipole field geometry to
match the 2.6 kG mean field and the 149 G longitudinal field for the
visible hemisphere of the star, we find that $i + \beta =
83.5^\circ$. However, in this case we have an additional constraint
from the \ion{He}{1} polarization.

If the photospheric magnetic field on TW Hya is dipolar, then the
magnetic poles must be nearly perpendicular to the line of sight at
all rotational phases (see preceding paragraph). For such a magnetic
geometry, material accreted from the corotation radius ($6.3\ R_*$,
see Johns--Krull \& Valenti 2001) will land within $13^\circ$ of the
magnetic pole. In this case, the longitudinal field of $\sim 1.7$ kG
measured in the accretion region using the \ion{He}{1} line is only
the small fraction projected onto the line of sight of the true
magnetic field in this region.  The minimum value of the true field
is $1.7\ {\rm kG} / {\rm cos(}83.5^\circ - 13^\circ{\rm )} = 5.1$
kG. Such a strong field in the \ion{He}{1} line has not been
observed in any of the previous studies now covering 11 CTTSs
(Johns--Krull et al. 1999a; Valenti \& Johns--Krull 2004; Symington
et al. 2005; Daou et al. 2006). We conclude that the surface
topology of the magnetic geometry on TW Hya (and likely all CTTSs)
is not a $pure$ dipole. It is likely that the field topology at the
stellar surface is dominated by small scale structure such as seen
on the Sun. However, the dipole component of the field will fall off
the least rapidly with distance from the star, so it may well be
that the interaction of the stellar field with the disk is governed
by a dipole geometry. Such a picture may explain the relatively
smooth, sinusoid like modulation of the field traced in the
\ion{He}{1} line of the CTTSs studied by Valenti and Johns--Krull
(2004).

Hartmann (1998) gives an expression for the truncation radius for an
assumed dipole field geometry (his equation 8.72) derived under the
assumption of spherical accretion.  As discussed by Bouvier et al.
(2006), this is an upper limit for a disk geometry.  If we assume
that the dipole component of the field on TW Hya is responsible for
our detection of \Bz\ $ = 149$ G on this star, we can use the
equation from Hartmann (1998) to estimate a truncation radius.  The
estimate again depends on the inclination of the dipole component of
the field with respect to our line of sight. If (as is generally the
case for the Sun) we assume the dipole component is aligned with the
rotation axis, assuming $i = 28^\circ$ gives the strongest possible
dipole component consistent with our \Bz\ detection and this
correponds to an equatorial field strength of 260 G. Putting this
into equation (8.72) from Hartmann (1998) yields a truncation radius
of 3 $R_*$ which is significantly less than the co-rotation radius
of 6.3 $R_*$.  Eisner et al. (2006) analyze K-band interferometry
observations of TW Hya from Keck, along with previous K-band veiling
and NIR photometric measurements, to conclude that the inner radius
of the optically thin disk is around 0.06 AU, which corresponds to
13 $R_*$.  While there are additional techniques that could shed light
on the exact location of this inner truncation radius (e.g. linear
polarimetry is a powerful technique which may help in this regard as
discussed in Vink et al. 2005), the current data suggests that the
system is not in equilibrium
as assumed by magnetospheric accretion theories if the dipole
component of the field alone is responsible for truncating the
accretion disk. There are certainly higher order contributions to
the total field at the surface of TW Hya which will contribute some
field strength at the co-rotation radius, but just how much depends
on the detailed field geometry. We suggest more work is needed to
verify whether the fields on TTSs really are strong enough to
truncate disks around these stars near the co-rotation radius for
realistic magnetic field geometries.

As noted in \S 3.2, the level of polarization in the \ion{Ca}{2}
$8498$ \AA\ line is much less than that detected in the \ion{He}{1}
line. The IRT lines of \ion{Ca}{2} likely have contributions from
different regions in CTTSs.  These lines sometimes show only broad
components (BC), narrow components (NC), or a mixture of the two
(e.g. Alencar \& Basri 2000).  It is generally accepted that the BC
of the IRT originates in the accretion and/or wind flows associated
with CTTSs, as this component is not observed in NTTSs.  However,
the origin of the NC of the IRT lines is not completely clear.  The
\ion{Ca}{2} $8498$ \AA\ line of TW Hya is dominated by a NC inside a
photospheric absorption line.  A chromospheric origin for the NC of
the IRT lines in both NTTSs and CTTSs was proposed by Hamann and
Persson (1992).  Batalha and Basri (1993) and Batalha et al.  (1996)
echo this idea, though they further suggest that accretion activity
in CTTSs can enhance the chromospheric emission.  They suggest that
this is due to the reprocessing of radiation produced in accretion
shocks as the accreting material hits the stellar atmosphere.
Another possible scenario results since the accretion occurs along
stellar magnetic field lines: the process may launch Alfv\'en waves
in the magnetosphere which deposits energy in the chromosphere,
heating it beyond what the star alone would do.  Observationally,
there is almost certainly a chromospheric contribution to this line
in TW Hya and the NC of other CTTSs given the persistent appearance
of NC emission in NTTSs.  For example, Batalha et al. (1996)
measured equivalent widths (EWs) of the \ion{Ca}{2} $8498$ \AA\ line
for 3 K7 NTTSs, finding values which range from $0.53$ \AA\ to
$0.59$ \AA.  These values are weaker than our EW measurements of TW
Hya ($0.69 - 0.98$ \AA), suggesting a contribution to the line in TW
Hya from more than just pure chromospheric emission.  This mixture
of sources can explain the relatively weak polarization observed in
the \ion{Ca}{2} line relative to the \ion{He}{1} line.  We generally
expect the magnetic field in the stellar chromosphere to show the
same behavior as that traced by the photospheric absorption lines.
In particular, the weak photospheric polarization suggests there
should be comparably weak polarization in the IRT lines, and we
further expect the implied field to be of the same polarity.  The
field detected in the \ion{Ca}{2} line is of opposite polarity to
that observed in the photosphere, implying that the polarization in
the enhanced portion of the IRT emission is more polarized than
implied by the reported value of \Bz\ in this line. The polarity of
the average field in the \ion{Ca}{2} region is the same as that in
the \ion{He}{1} region on all nights, suggesting that their emission
is related. Whether this implies actual formation of (some of) the
IRT NC emission in the accretion shock or formation in nearby field
regions of the same polarity is not clear.

This dilution of the \ion{Ca}{2} polarization by chromospheric
emission is not expected to be an issue for the \ion{He}{1} line.
The \ion{He}{1} $5876$ \AA\ line of TW Hya is dominated by NC
emission as well; however, the contribution to this from a pure
stellar chromosphere is likely quite small. For example, Batalha et
al. (1996) find the EW of the \ion{He}{1} $\lambda 5876$ for the
same 3 K7 NTTSs to range between $0.00 - 0.09$ \AA, while for TW Hya
we find values from $2.5 - 4.3$ \AA. Besides chromospheric
activities, both magnetospheric infall and a hot wind could
potentially give rise to the \ion{He}{1} emission. They are believed
to be responsible for the blue-shifted or red-shifted broad
components of the $5876$ \AA\ line profiles in spectra of a number
of T Tauri stars (e.g., see Baristain et al. 2001). In the case of
TW Hya, an obvious BC is not seen. Thus, the \ion{He}{1} line in TW
Hya, and indeed in most CTTSs, is likely dominated by emission from
the accretion shock. In the extreme case of Batalha et al. (1996),
the collective non-shock contributions only account for 4\% (0.09
\AA / 2.5 \AA) of the \ion{He}{1} emission, which makes our estimate
that 16\% of \ion{Ca}{2} emission is from accretion regions only
adjusted down by 0.6\% at most.

We did not detect any polarization in the H$\alpha$ line forming
region. H$\alpha$ emission likely results from several zones in the
TW Hya system including the accretion shock, magnetospheric
accretion flow, and the stellar and/or disk wind flowing away from
the star.  Thus, the line forms over a large volume where the mean
magnetic field is likely relatively weak and the direction of the
field is not constant.  Indeed, there are regions in the
magnetosphere (if viewed nearly pole-on) where the field directions
are reversed.  This mixture of polarities will reduce the separation
of LCP and RCP line profiles.  Depending on where the majority of
the line forms, we expect the local field strength to be quite weak.
For example, assuming a face-on disk for TW Hya and a dipolar-like
magnetospheric accretion flow with the magnetic axis close to the
stellar rotation axis, the dipole component of the field at the
corotation radius will be equal to the stellar value of that field
multiplied by $(R_*/R_{co})^3$.  Taking our estimate above of 260 G
for the equatorial value of the dipole component of the field at the
stellar surface and $R_{co} = 6.3 R_*$ yields a field of about 1 G
at a corotation radius where the magnetospheric flow may originate.
This example simply illustrates that the field strength throughout a
large portion of the H$\alpha$ line forming region may be quite
weak, so our lack of a detection in this line may well be expected.

\acknowledgements
We would like to thank the referee, J. Vink, for many useful comments and
suggestions for improving the original manuscript.
CMJ-K and HY would like to acknowledge partial support from the NASA Origins
of Solar Systems program through grant numbers NAG5-13103 and NNG06GD85G
 made to Rice University.

\clearpage


\clearpage

%
 \begin{table}
\begin{minipage}{10cm}
    \label{obs}
    \caption {Observations and Results}
    \begin{tabular}[h]{cccccc}
    \tableline\tableline

   \  UT Date     & UT Time & $N_{\rm{exp}}$ & Total Exposure   & $B_{\rm{z}}$\footnote[1]{Photospheric field strength.}(G) & $\sigma$ \\[+5pt]
   \              &         &                &    Time(s)       &                                                           &          \\[+5pt]
    \tableline

     1999 Apr 21  & 03:54   &    2           & 4300                    &  66 & 40  \\[+5pt]
     1999 Apr 22  & 04:12   &    2           & 4700                    &  54 & 56  \\[+5pt]
     1999 Apr 23  & 03:44   &    2           & 4700                    &  84 & 32  \\[+5pt]
     1999 Apr 24  & 03:38   &    2           & 4700                    &  82 & 61  \\[+5pt]
     1999 Apr 25  & 04:48   &    2           & 4700                    &  47 & 48  \\[+5pt]
     1999 Apr 26  & 03:37   &    2           & 4700                    & 149 & 33  \\[+5pt]

    \tableline
  \end{tabular}
\end{minipage}

\end{table}

\begin{table}

  \caption{Photospheric Field Measurements.}
  \centering
 \begin{center}  \label{linedata}
 \leavevmode
  \footnotesize
   \begin{tabular}[h]{ccccccccc}
    \tableline\tableline
 \ Species  &$\lambda(\rm{\mbox{\AA}})$& $\geff$        &  Apr 21         &  Apr 22       &  Apr 23       & Apr 24         &  Apr 25        & Apr 26       \\[+5pt]
  \tableline
\ion{Ca}{1} &   6166.4          &   0.50         &  186 $\pm$ 393  &  -78$\pm$ 532 &  258$\pm$ 343 &  -450$\pm$ 586 &   76$\pm$ 316  &  -448$\pm$ 321\\[+5pt]
\ion{Fe}{1} &   6180.2          &   0.64         &  150 $\pm$ 350  & -269$\pm$ 418 &  -79$\pm$ 307 &  -338$\pm$ 516 &  -20$\pm$ 349  &   280$\pm$ 247\\[+5pt]
\ion{Fe}{1} &   6200.3          &   1.51         &  173 $\pm$ 150  &  -79$\pm$ 158 &   52$\pm$ 102 &    74$\pm$ 226 &  238$\pm$ 143  &   113$\pm$  84\\[+5pt]
\ion{V}{1}  &   6213.4          &   2.00         &   -6 $\pm$ 111  &   30$\pm$ 159 &  108$\pm$ 113 &   145$\pm$ 224 & -277$\pm$ 156  &    92$\pm$  86\\[+5pt]
\ion{Fe}{1} &   6322.6          &   1.51         &   47 $\pm$ 133  &   14$\pm$ 328 &   31$\pm$ 117 &  -110$\pm$ 227 &  283$\pm$ 235  &    79$\pm$ 117\\[+5pt]
\ion{Fe}{1} &   6330.8          &   1.22         &   12 $\pm$ 199  &  233$\pm$ 258 &   75$\pm$ 148 &    74$\pm$ 281 &  -41$\pm$ 194  &    13$\pm$ 123\\[+5pt]
\ion{Fe}{1} &   6335.3          &   1.16         &  222 $\pm$ 161  &  244$\pm$ 245 &  190$\pm$ 118 &   540$\pm$ 273 &  162$\pm$ 196  &   280$\pm$ 138\\[+5pt]
\ion{Fe}{1} &   6336.8          &   2.00         &   76 $\pm$ 144  &  107$\pm$ 237 &  122$\pm$  92 &   245$\pm$ 199 &  114$\pm$ 140  &   157$\pm$  89\\[+5pt]
\ion{Ti}{1} &   6359.8          &   1.20         &   34 $\pm$ 145  &  197$\pm$ 231 &  110$\pm$ 112 &    71$\pm$ 217 &  317$\pm$ 148  &   359$\pm$ 111\\[+5pt]
\ion{Al}{1} &   6696.0          &   1.16         &   50 $\pm$ 258  &  101$\pm$ 337 &  -39$\pm$ 304 &   123$\pm$ 462 &  -11$\pm$ 372  &   111$\pm$ 267\\[+5pt]
\ion{Fe}{1} &   8468.4          &   2.50         &   43 $\pm$  66  &   22$\pm$ 101 &   47$\pm$  57 &    50$\pm$ 107 &  -53$\pm$  92  &   181$\pm$  74\\[+5pt]
\ion{Fe}{1} &   8757.1          &   1.50         &  142 $\pm$ 168  &  106$\pm$ 149 &  131$\pm$ 103 &    35$\pm$ 170 & -112$\pm$ 181  &   156$\pm$ 140\\[+5pt]
\tableline
 Telluric   &   8139.5          &   0.90         &  -33 $\pm$  88  &   24$\pm$  83 &   55$\pm$  95 &  -153$\pm$  89 &  -62$\pm$  65  &   -75$\pm$ 243\\[+5pt]
 Telluric   &   8140.5          &   0.90         &   17 $\pm$  48  &  124$\pm$  51 &   14$\pm$  55 &    78$\pm$  62 &   35$\pm$  76  &   -21$\pm$  72\\[+5pt]
 Telluric   &   8141.5          &   0.90         &  -33 $\pm$  45  &  -30$\pm$  64 &  -31$\pm$  44 &    21$\pm$  69 &  -78$\pm$  81  &    25$\pm$  57\\[+5pt]
 Telluric   &   8146.1          &   0.90         &  -10 $\pm$  69  &   -4$\pm$  71 &   44$\pm$  68 &   -38$\pm$  58 &  -21$\pm$  79  &    33$\pm$  61\\[+5pt]
 Telluric   &   8147.0          &   0.90         &   64 $\pm$  68  &   58$\pm$  82 &   34$\pm$  66 &    -8$\pm$  68 &  -27$\pm$  83  &    27$\pm$  69\\[+5pt]
 Telluric   &   8158.0          &   0.90         &   -3 $\pm$  37  &    3$\pm$  47 &  -10$\pm$  47 &   -82$\pm$  66 &  -37$\pm$  73  &   -17$\pm$  51\\[+5pt]
\tableline
\ion{He}{2} &   5876.           &   1.00         & -1806$\pm$ 114  &-1506$\pm$ 112 &-1790$\pm$ 149 & -1583$\pm$ 159 &-1471$\pm$ 143  & -1776$\pm$  96\\[+5pt]
\ion{Ca}{2} &   8498.           &   1.07         &  -212$\pm$  57  & -289$\pm$  59 & -323$\pm$  36 &  -392$\pm$  62 & -402$\pm$  64  &  -180$\pm$  34\\[+5pt]
 \tableline
 \end{tabular}

 \end{center}

\end{table}

\clearpage

 \clearpage

\clearpage

\clearpage
\begin{figure}[ht]
  \begin{center}
  \includegraphics[scale=0.65]{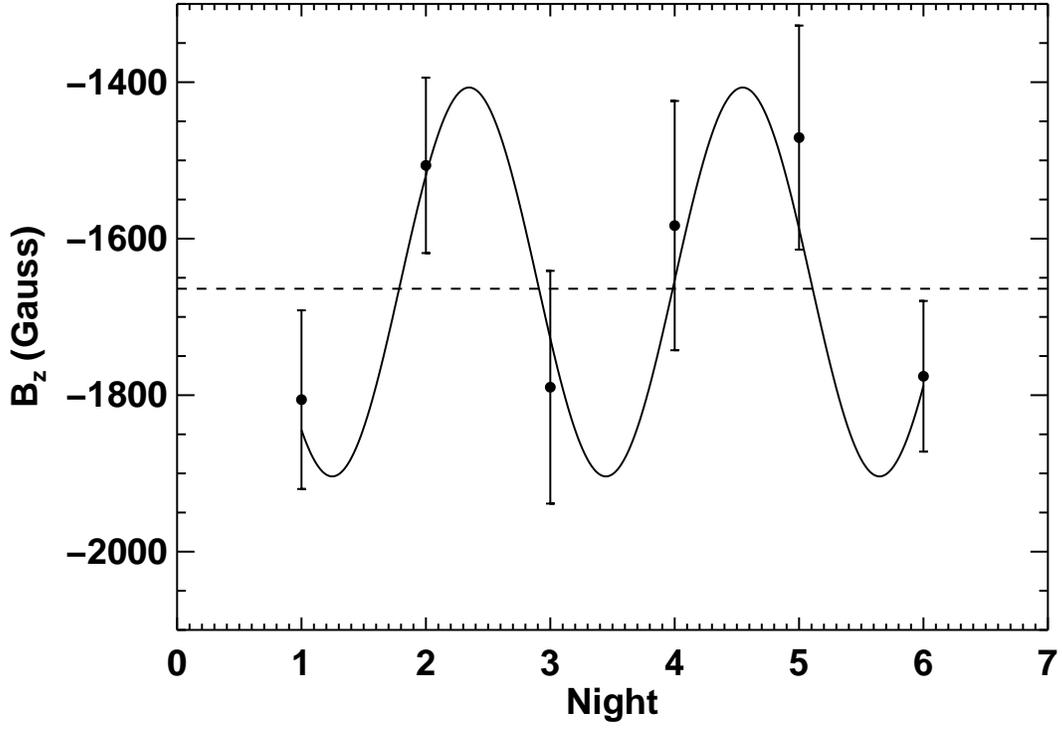}
  \end{center}
  \caption{Time series of the longitudinal magnetic field measured by \ion{He}{1}
  $\lambda5876$ line. The solid curve is a sine wave fit with a period of 2.2
  days. The dash line is a best fit model with no periodicity.}
\label{he_period}
\end{figure}

\clearpage
\begin{figure}[ht]
  \begin{center}
  \includegraphics[scale=0.65]{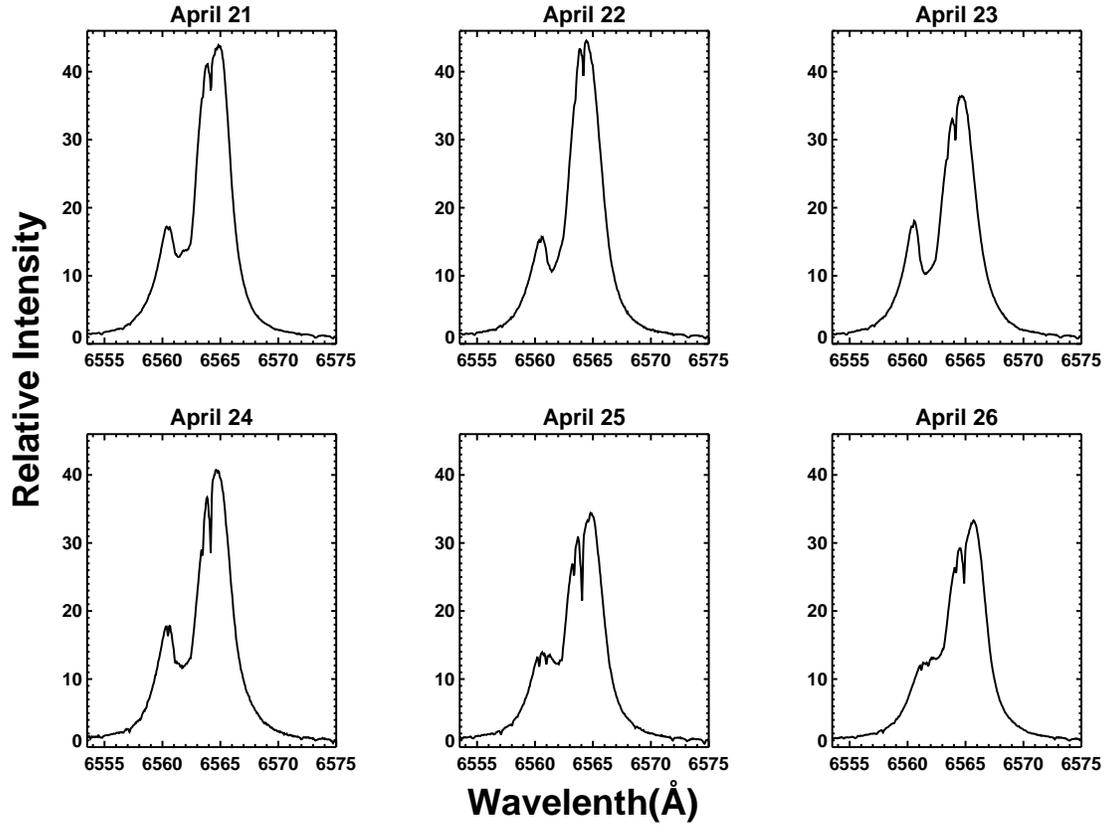}
  \end{center}
  \caption{$\rm{H\alpha}$ line profiles from six consecutive nights.}
\label{halpha}
\end{figure}

\end{document}